\documentclass[12pt,draftclsnofoot,journal,onecolumn]{IEEEtran}

\usepackage[english]{babel}
\usepackage{amsmath}
\usepackage{amssymb}
\usepackage{esint}
\usepackage{bbm}
\usepackage{color,graphicx}
\usepackage{mathrsfs}
\usepackage{subfigure}

\newtheorem{theorem}{Theorem}[section]         
\newtheorem{lemma}[theorem]{Lemma}             
\newtheorem{corollary}[theorem]{Corollary}     
\newtheorem{definition}[theorem]{Definition}   
\newtheorem{remark}[theorem]{Remark}           
\newtheorem{proposition}[theorem]{Proposition} 

\renewcommand{\QED}{\QEDopen}

\renewcommand{\vec}{\boldsymbol}								
\newcommand{\set}[1]{{\cal{#1}}}								
\newcommand{\bb}[1]{{\mathbbm{#1}}}							
\newcommand{\N}{\mathbb N}

\newcommand{\eps}{{\varepsilon}}

\newcommand{\Rro}{R_{\mathrm{\overrightarrow{\mathrm{R}1}}}}
\newcommand{\Rrt}{R_{\mathrm{\overrightarrow{\mathrm{R}2}}}}

\newcommand{\Ror}{R_{\mathrm{\overrightarrow{1\mathrm{R}}}}}
\newcommand{\Rtr}{R_{\mathrm{\overrightarrow{2\mathrm{R}}}}}

\newcommand{\nbc}{{n}_{\mathrm{BC}}}
\newcommand{\nmac}{{n}_{\mathrm{MAC}}}

\newcommand{\sCmac}{\set{C}_{\mathrm{MAC}}}
\newcommand{\sCbc}{\set{C}_{\mathrm{BC}}}
\newcommand{\sCbcmax}{\set{C}_{\mathrm{BC,max}}}
\newcommand{\sCbcav}{\set{C}_{\mathrm{BC,av}}}
\newcommand{\sRbrc}{\set{R}_{\mathrm{BRC}}}

\title{Broadcast Capacity Region of Two-Phase Bidirectional Relaying\thanks{This work was partly supported by 
the DFG via projects \mbox{Bj 57/1-1} and BO 1734/7-1 and by the German Ministry for Education and 
Research (BMBF) under Grant 01BU0680.}}
\IEEEoverridecommandlockouts

\author{ \authorblockN{
		Tobias J. Oechtering\authorrefmark{1}~\IEEEmembership{Student Member,~IEEE,}\\
		Igor Bjelakovi\'c\authorrefmark{1}\authorrefmark{2},\\
		Clemens Schnurr\authorrefmark{3}~\IEEEmembership{Student Member,~IEEE,} and\\
		Holger Boche\authorrefmark{1}\authorrefmark{2}~\IEEEmembership{Senior Member,~IEEE}\\} 
\authorblockA{\small
    \authorrefmark{1} 		Technical University of Berlin, Heinrich-Hertz-Chair for Mobile Communications,
    Einsteinufer 25, 10587 Berlin, Germany.\\
    \authorrefmark{2} 		Technical University of Berlin, Institut f\"ur Mathematik,
    Stra\ss e des 17. Juni 136, 10623 Berlin, Germany.\\
\authorrefmark{3} Fraunhofer German-Sino Lab for Mobile Communications,
    Einsteinufer 37, 10587 Berlin, Germany.\\}
    }
    
\begin{document}
\maketitle
\begin{abstract} In a three-node network a half-duplex relay node enables bidirectional communication between two nodes with a spectral efficient two phase protocol. In the first phase, two nodes transmit their message to the relay node, which decodes the messages and broadcast a re-encoded composition in the second phase. In this work we determine the capacity region of the broadcast phase. In this scenario each receiving node has perfect information about the message that is intended for the other node. The resulting set of achievable rates of the two-phase bidirectional relaying includes the region which can be achieved by applying XOR on the decoded messages at the relay node. We also prove the strong converse for the maximum error probability and show that this implies that the $[\eps_1,\eps_2]$-capacity region defined with respect to the average error probability is constant for small values of error parameters $\eps_1$, $\eps_2$.
\end{abstract}


\section{Introduction}
Future wireless systems should offer connectivity almost everywhere. This objective represents an ambitiously engineering challenge in scenarios where the direct link between two nodes does not have the desired quality, e.g. due to shadowing or distance. On that score, multi-hop communication for coverage extension and meshed network architectures are currently discussed or scheduled in all wireless networks standards of the next generation. Therefore, the relay channel experiences a revival recently. The problem was introduced by van der Meulen in \cite{Meulen} in the early seventies. A few years later, Cover and El Gamal obtained the capacities of the physically degraded and reversely degraded relay channels and upper and lower bounds on the capacity of the general relay channel in \cite{CovGam}. The general problem is still unsolved. Fundamental insights about the general problem and recent development can be  found in \cite{KGG05csac} and references therein.

We consider a three-node network where one node acts as a relay to enable the bidirectional communication between two other nodes. The two-way communication problem without a relay node was introduced by Shannon in \cite{Shan61} in 1961 already.  Therein, he obtained the capacity region for the average error for the restricted two-way channel, i.e. a feedback between the two nodes is not allowed. Nowadays, this is regarded as the first network information theory problem. 

In information theory it is often assumed that the nodes can transmit and receive at the same time, i.e. full-duplex nodes. This assumption is in wireless communication hard to fulfill, since it is practically difficult to isolate a simultaneously received and transmitted signal using the same frequency sufficiently. Therefore, in this work we assume half-duplex nodes. As a natural consequence of this assumption is that relay communication is performed in phases. Often the relay communication should be integrated in existing infrastructures and most protocol proposals base usually on orthogonal components which require exclusive resources for each link. As a consequence they suffer from an inherent loss in spectral efficiency. This loss can be significantly reduced if bidirectional relay communication is desired. Because then the communication can be efficiently performed in two phases. In the first phase, the multiple access phase (MAC), the information is transmitted to the relay node. In the succeeding broadcast phase (BC), the relay node forwards the information to its destinations. In \cite{RW05ses} and \cite{OB06oraf}, where Gaussian channels are considered, the relay performs superposition encoding in the second phase. The knowledge of the first phase allows the receiving nodes to perform interference cancellation before decoding so that effectively we achieve interference-free transmission in the second phase. Another interesting approach \cite{WuChouKung05}, \cite{FBW06nca} is based on the network coding principle \cite{Ahli}, \cite{NOWYeung} where the relay node performs an XOR operation on the decoded bit streams. But since network coding  is  originally a multi-terminal source coding problem, such an approach operates on the decoded data and therefore does not deal with channel coding aspects. 

Because of our practical motivation, we apply time-division to separate the bidirectional relay communication into two phases. The optimal coding strategy and capacity region of the general multiple access channel is known. In this work, we present the optimal broadcast coding strategy of the two-phase bidirectional relay channel based on classical channel coding. It shows that all rate pairs in the capacity region can be achieved using an auxiliary random variable taking two values, i.e. we achieve the capacity region by the principle of time-sharing. Thereby, we see an interesting connection to a joint source and channel coding approach for the broadcast channel based on Slepian-Wolf coding \cite{Tuncel}.

In a multi-terminal system the average and maximal error capacity region can be different, even in the case of asymptotically vanishing errors as is shown by Dueck in \cite{dueck-counter}.
While for single-user channels it is of no importance whether we use vanishing average or maximal probabilities of error in the definition  of achievable rates, the choice of the error criterion makes a big difference if we pass to the consideration of the strong converses for one-way channels. Indeed Ahlswede demonstrated in \cite{ahl-counter} that the strong converse does not hold for the compound channels if we use the average probability of error for the definition of $\eps$-achievable rates but it is well known that the strong converse is valid if we use maximal error probabilities as was shown by Wolfowitz \cite{wolfo-compound}. For these reasons, we will pay a lot of attention to the consideration of the maximal and average error probabilities and the relation between them in the main part of the paper and in the proofs.

The paper is organized as follows: In the following two subsections we present the two-phase bidirectional relay model, which describes the context of the bidirectional broadcast channel and after that we briefly restate the MAC capacity region for completeness.
In Section \ref{sec:BCcap} we prove a coding theorem and a weak converse for the maximum error probability.
The proof shows that the capacity region is independent of whether we use asymptotically vanishing average or maximum probability of error. In Section \ref{sec:StrongConv} we prove the strong converse for the maximum error probability using the Blowing-up Lemma \cite{AGK}. Finally, from this we can deduce that the $[\eps_1,\eps_2]$-capacity region in terms of average probability of error is constant for all $[\eps_1,\eps_2]\in (0,\frac{1}{2})\times (0,\frac{1}{4})$ or $\in(0,\frac{1}{4})\times (0,\frac{1}{2}) $ and equals the $[\eps_1,\eps_2]$-capacity region defined with respect to maximum error probability in that range of values of $[\eps_1,\eps_2]$. 
Based on the capacity regions of the two phases the time-division between MAC and BC phase can be optimized. This gives us the largest achievable rate region for the finite alphabet discrete memoryless bidirectional relay channel under the simplification of time-division into two phases, which will be discussed in Section \ref{sec:discussion} by means of a binary channel example.


\subsection{Two Phase Bidirectional Relay Channel}
We consider a three-node network with two message sets $\mathcal{W}_1$ and $\mathcal{W}_2$. In our bidirectional channel we want the messages $w_1\in\mathcal{W}_1$ located at node 1 and the message $w_2\in\mathcal{W}_2$ located at node 2 to be known at node 2 and node 1, respectively. We assume that there is no direct channel between node 1 and 2. Therefore, node 1 and 2 need the support of a relay node R. 

\begin{figure}
\begin{center}	
	\includegraphics[width=8cm]{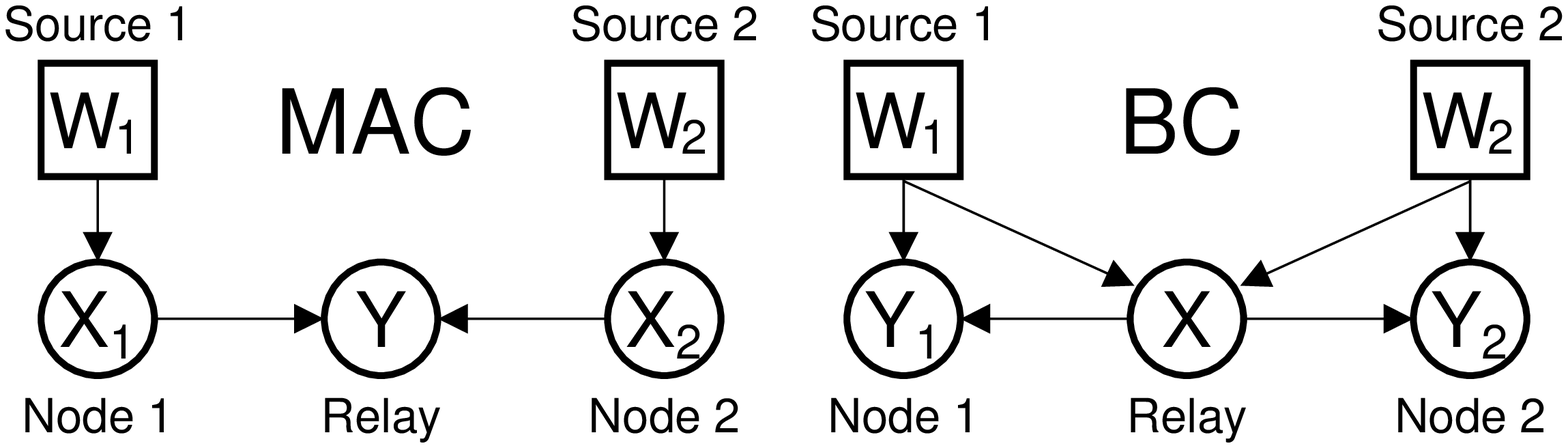}
\end{center}
\caption{Multiple access (MAC) and broadcast (BC) phase of the time division bidirectional relay channel.}
\label{fig:model}
\end{figure}

We simplify the problem by assuming an a priori separation of the communication into two phases. Furthermore, we do not allow cooperation between the encoders at node 1 and node 2. Otherwise, a transmitted symbol could depend on previously received symbols. For a two-way channel this is known as a restricted two-way channel. With this simplification we end up with a multiple access phase, where node 1 and 2 transmit messages $w_1$ and $w_2$ to the relay node, and a broadcast phase, where the relay forwards the messages to node 2 and 1, respectively. We look at the two phases separately. After that we will briefly consider the optimal time-division between the two phases.

In the multiple access phase (MAC) we have a classical multiple access channel, where the optimal coding strategy and capacity region $\sCmac$ is known \cite{AhliMAC}, \cite{LiaoMAC}. We will restate the capacity region in the next subsection.  Thereby,  let $\Ror$ and $\Rtr$  denote the achievable rates between  node 1 and 2 and the relay node in the MAC phase.

For the broadcast phase (BC), we assume that the relay node has successfully decoded the messages $w_1$ and $w_2$ in the multiple access phase. From the union bound we know that the error probability of the two-phase protocol is at most the sum of the error probability of each phase. Therefore, an error-free MAC phase is reasonable if we assume rates within the MAC capacity region and a sufficient coding length. From this we have a broadcast channel where the message $w_1$ is known at node 1 and the relay node  and the message $w_2$ is known at node 2 and the relay node, as depicted in Figure \ref{fig:model}. Thereby, let $x_1$, $x_2$ and $x$ denote the input and $y_1$, $y_2$ and $y$ the output symbols of node 1, node 2, and the relay node, respectively. Furthermore, let $\Rro$ and $\Rrt$ denote the achievable rates between the relay node and node 1 and 2 in the BC phase.

The mission of the relay node is to broadcast a message to node 1  and 2 which allows them to recover the unknown source. This means that node 1 wants to recover message $w_2$ and node 2 wants to recover message $w_1$. 
We will present an information theoretic optimal coding strategy and the capacity region of the bidirectional broadcast channel in Section  \ref{sec:BCcap}.


\subsection{Capacity Region of Multiple Access Phase}
In this subsection, we restate the capacity region of the multiple access channel, which was found by Ahlswede \cite{AhliMAC} and Liao \cite{LiaoMAC} and is part of any textbook on multiuser information theory, e.g. \cite{CsiszarKoerner}.

\begin{definition} A {\it discrete memoryless multiple access channel} is the family $\{p^{(n)}:\set{X}_{1}^{n}\times \set{X}_{2}^{n}\to\set{Y}^{n}\}_{n\in\N}$ with finite input alphabets $\set{X}_{k}$, $k=1,2$, and the finite output alphabet $\set{Y}$ where the probability transition functions are given by $p^{(n)}(y^n|x_1^n,x_2^n):=\prod_{i=1}^{n}p(y_i|x_{i1},x_{i2})$ for a given probability transition function $\{p(y|x_1,x_2)\}_{x_1\in \set{X}_1,x_2\in\set{X}_2,y\in\set{Y}}$.
\end{definition}


\begin{theorem}
The capacity region $\sCmac$ of the memoryless multiple access channel is the set of all rate pairs $[\Ror,\Rtr]$ satisfying 
\begin{align*}
\Ror&\leq I(X_1;Y|X_2,U),\\ 
\Rtr&\leq I(X_2;Y|X_1,U), \text{ and} \\
\Ror+\Rtr&\leq I(X_1,X_2;Y|U),
\end{align*}
for random variables $[U,X_1,X_2,Y]$ with values in $ \set{U}\times \set{X}_1\times \set{X}_2\times \set{Y}$ and joint distribution $\{q(u)q_1(x_1|u)q_2(x_2|u)p(y|x_1,x_2)\}_{u\in\set{U},x_1\in\set{X}_1,x_2\in\set{X}_2,y\in\set{Y}}$. Furthermore, the range $\set{U}$ of the auxiliary random variable $U$ has a cardinality bounded by $|\set{U}|\leq 2$.
\end{theorem}


\section{Capacity Region of Broadcast Phase}
\label{sec:BCcap}

In this section we present our main result, the capacity region of a broadcast channel where the receiving nodes have perfect knowledge about the message which should be transmitted to the other node. The capacity region can be achieved by classical channel coding principles.
First we need to introduce some standard notation.

\begin{definition} Let $\set{X}$ and $\set{Y}_k$, $k=1,2$, be finite sets. A {\it discrete memoryless broadcast channel} is defined by a family $\{p^{(n)}:\set{X}^n\to \set{Y}_1^n\times \set{Y}_2^n   \}_{n\in\N}$ of probability transition functions given by $p^{(n)}(y_1^n,y_2^n|x^n):=\prod_{i=1}^{n}p(y_{i1},y_{i2}|x_i)$ for a probability transition function $p:\set{X}\to\set{Y}_1\times \set{Y}_2 $, i.e. $\{p(y_1,y_2|x)\}_{x\in\set{X},y_1\in\set{Y}_1,y_2\in\set{Y}_2}$ is a stochastic matrix. 
\end{definition}
In what follows we will suppress the super-index $n$ in the definition of the $n$-th extension of the channel $p$, i.e. we will write simply $p$ instead of $p^{(n)}$. This should cause no confusion since it will be always clear from the context which block length is under consideration. In addition, we will use the abbreviation $\set{V}:=\set{W}_1\times \set{W}_2$, where
$\set{W}_1$ and $\set{W}_2$ denote the message sets.
\begin{definition}
A $(M_1^{(n)},M_2^{(n)},n)$-code for the {\it bidirectional broadcast channel} 
consists of one encoder at the relay node
$$x^n:\set{V}\to\set{X}^n,$$
and a decoder at node one and two
\begin{align*}
g_1:\;\set{Y}_1^n\times\set{W}_1\to\set{W}_2\cup\{0\},\\
  g_2:\;\set{Y}_2^n\times\set{W}_2\to\set{W}_1\cup\{0\}.
\end{align*}
The element $0$ in the definition of the decoders is included for convenience only and plays the role of an erasure symbol.
\end{definition}

When the relay node sends the message $v=[w_1,w_2]$, the receiver of node one is in error if $g_1(Y^n_1,w_1)\neq w_2$. The probability of this event is denoted by
$$
\lambda_1(v):=\bb{P}[
g_1(Y^n_1,w_1)\neq w_2\, |\, x^n(v) \text{ has been sent}].
$$
Accordingly, we denote the probability that the receiver of node two is in error by
$$
\lambda_2(v):=\bb{P}[
g_2(Y^n_2,w_2)\neq w_1\, |\, x^n(v) \text{ has been sent}].
$$ 
Hereby, $Y_1^n$ and $Y_2^n$ denote the random outputs at nodes $1$ and $2$ given that the sequence $x^n(v)$ has been sent down the channel.
This allows us to introduce the notation for the maximum and average probability of error for the $k$-th node
\begin{align*}
\lambda_k^{(n)}:=
\max\limits_{v\in\set{V}}\lambda_k(v),\qquad
\mu_k^{(n)}:=\tfrac{1}{|\set{V}|}{\textstyle\sum\limits_{v\in\set{V}}}
\lambda_k(v).
\end{align*}

\begin{definition}
A rate pair $[\Rrt,\Rro]$ is said to be {\it achievable} for the bidirectional broadcast channel if for any $\delta >0$ there is an $n(\delta)\in\N$ and a sequence of $(M_1^{(n)},M_2^{(n)},n)$-codes such that for all $n\ge n(\delta)$ we have $\frac{\log M_1^{(n)}}{n}\ge\Rrt-\delta$ and $\frac{\log{M_2^{(n)}}}{n}\ge \Rro -\delta$ while $\lambda_1^{(n)},\lambda_2^{(n)}\to 0$ when $n\to\infty$. The set of all achievable rate pairs is the {\it capacity region} of the bidirectional broadcast channel and is denoted by $\sCbc $.
\end{definition}
\begin{remark}
Achievable rate pairs and a capacity region can be also defined for average probability of error. 
\end{remark}

\begin{theorem}
\label{theorem:capacity}
The capacity region $\sCbc$ of the bidirectional memoryless broadcast channel is the set of all rate pairs $[\Rrt,\Rro]$ satisfying
\begin{equation}
\begin{split}
	\Rrt\leq I(X;Y_2|U), \\
	\Rro\leq I(X;Y_1|U), \label{eq:RATEwithTSH}
\end{split}
\end{equation}
for random variables $[U,X,Y_1,Y_2]$ with values in $\set{U}\times \set{X}\times \set{Y}_1\times \set{Y}_2$ and joint probability distribution $\{q_1(u)q_2(x|u)p(y_1,y_2|x)\}_{u\in\set{U},x\in \set{X},y_1\in\set{Y}_1,y_2\in\set{Y}_2}$. The cardinality of the range of $U$ can be bounded by $|\set{U}|\leq 2$.
\end{theorem}

The theorem is proved in the following three subsections. In the first subsection we prove the achievability, i.e. a coding theorem. We prove a weak converse with respect to the maximum probability of error in the second subsection. Then the theorem is proved with the third subsection where we show that a cardinality of two is enough for the range of the auxiliary random variable. 


\subsection{Proof of Achievability}
Here, we adapt the random coding proof for the degraded broadcast channel of \cite{Beg73} to 
our context. First, we prove the achievability of all rate pairs $[\Rrt,\Rro]$ satisfying
\begin{equation}
	\Rrt\le I(X;Y_2), \quad	
	\Rro\le I(X;Y_1), 
	\label{eq:RATEwithoutTSH}
	\end{equation}
for some probability function $p(x)p(y_1,y_2|x)$. 
Then, we extend this to prove that all points in the closure of the convex hull of \eqref{eq:RATEwithoutTSH} are achievable, which we will see is exactly the region stated in  Theorem \ref{theorem:capacity}.

\subsubsection{ Random codebook generation}
We generate $M_1^{(n)} M_2^{(n)}$ 
independent codewords $X^n(v)$, $v=[w_1,w_2]$ of length $n$
with $M_1^{(n)}:=2^{\lfloor n\Rrt\rfloor}$ and $M_2^{(n)}:=2^{\lfloor n\Rro\rfloor}$ according to $\prod_{i=1}^n p(x_i)$.

\subsubsection{ Encoding} 
To send the pair $v=[w_1,w_2]$ with $w_k\in\set{W}_k$, $k=1,2$, the relay sends the corresponding codeword
$x^n(v)$.

\subsubsection{ Decoding}
The receiving nodes use typical set decoding. First, we characterize the decoding sets. For the decoder at node $k=1,2$ let
\begin{equation*}
I(x^n;y^n_k):=\tfrac{1}{n}\log_2\frac{p(y^n_k|x^n)}{p(y^n_k)}
\end{equation*}
with average mutual information $I(X;Y_k):=\bb{E}_{x^n,y^n_k}[I(x^n;y^n_k)]$.
 This gives the decoding set
\begin{equation*}
\set{S}(y^n_k):=\left\{{x^n}\in\set{X}^n:I(x^n;y^n_k)\geq\tfrac{R_{\overrightarrow{\mathrm{R k}}}+I(X;Y_k)}{2}\right\}
\end{equation*}
and indicator function
$$
d(x^n,y^n_k):=\begin{cases}
1, & \text{ if } x^n\notin \set{S}(y^n_k)\\
0, & \text{ otherwise.}
\end{cases}
$$
When $x^n(v)$ with $v=[w_1,w_2]$ has been sent, and $y^n_1$ and $y^n_2$ have been received
we say that the decoder at node $k$ makes an error if either $x^n(v)$  is not in $\set{S}(y^n_k)$ (occurring with probability $P_{e,k}^{(1)}(v)$) or if 
at node one $x^n(w_1,\hat{w}_2)$ with 
$\hat{w}_2\neq w_2$ is in $\set{S}(y^n_1)$  
or at node two $x^n(\hat{w}_1,w_2)$ with  
$\hat{w}_1\neq w_1$ is in $\set{S}(y^n_2)$  (occurring with $P_{e,k}^{(2)}(v)$). If there is no or more than one codeword $x^n(w_1,\cdot)\in\set{S}(y^n_1)$ or $x^n(\cdot,w_2)\in\set{S}(y^n_2)$, the decoders map on the 
erasure symbol $0$.

\subsubsection{ Analysis of the probability of error} 
From the union bound we have $
\lambda_k(v)\leq 
P_{e,k}^{(1)}(v)+
P_{e,k}^{(2)}(v)
$
with 
$$P_{e,k}^{(1)}(v):={\textstyle \sum\limits_{y_k^n\in\set{Y}^n_k}}
p(y_k^n|x^n(v))\,d(x^n(v),y_k^n)\quad  \text{for } k=1,2$$ and 
\begin{equation*}
\begin{split}
P_{e,1}^{(2)}(v):=&{\textstyle\sum\limits_{y_1^n\in\set{Y}^n_1}}
p(y_1^n|x^n(v))\,{\textstyle\sum\limits_{\substack{\hat{w}_2=1\\\hat{w}_2\neq w_2}}^{|\set{W}_2|}}
\big(1-d(x^n(w_1,\hat{w}_2),y_1^n)\big),\\
P_{e,2}^{(2)}(v):=&
{\textstyle \sum\limits_{y_2^n\in\set{Y}^n_2}}
p(y_2^n|x^n(v))\,
{\textstyle\sum\limits_{\substack{\hat{w}_1=1\\\hat{w}_1\neq w_1}}^{|\set{W}_1|}}
\big(1-d(x^n(\hat{w}_1,w_2),y_2^n)\big).
\end{split}
\end{equation*}

For uniformly distributed messages $W_1$ and $W_2$ we define
$P_{e,k}^{(m)}:=\frac{1}{|\set{W}_1|\,|\set{W}_2|}\sum_{v\in\set{W}_1\times\set{W}_2}
P_{e,k}^{(m)}(v)$ for $m=1,2$
so that  $\mu_k^{(n)}\leq P_{e,k}^{(1)}+P_{e,k}^{(2)}$. Next, we average over 
all codebooks, i.e.
$
\bb{E}_{x^n}[\mu_k^{(n)}]
\leq\bb{E}_{x^n}[P_{e,k}^{(1)}+P_{e,k}^{(2)}]$. 

In the following, we  show that
if $R_{\overrightarrow{\mathrm{R} k}}\leq I(X,Y_k)-2\eps$ for any $\eps>0$, we have $\bb{E}_{x^n}[\mu_k]\rightarrow 0$ when $n\rightarrow \infty$. We have
\begin{equation*}
\begin{split}
\bb{E}_{x^n}[P_{e,k}^{(1)}]=&
\frac{1}{|\set{W}_1|\,|\set{W}_2|}\sum_{v\in\set{W}_1\times\set{W}_2}
\bb{E}_{x^n}[P_{e,k}^{(1)}(v)]\\
\underset{\text{fixed } v}{\overset{\text{for any}}{=}}&
\sum\limits_{y_k^n\in\set{Y}^n_k}
\bb{E}_{x^n}[p(y_k^n|x^n(v))\,d(x^n(v),y_k^n)]\\
=&\sum\limits_{y_k^n\in\set{Y}^n_k}\sum\limits_{x^n\in\set{X}^n}
p(x^n)p(y_k^n|x^n)\,d(x^n,y_k^n)\\
=&\bb{E}_{x^n,y_k^n}[d(x^n,y_k^n)]=\bb{P}[d(x^n,y_k^n)=1]\\
=&\bb{P}\left[
I(x^n;y_k^n)\leq\tfrac{R_{\overrightarrow{\mathrm{R}k}}+I(X;Y_k)}{2}
\right]\\
\leq&\bb{P}\left[
I(x^n;y_k^n)\leq I(X;Y_k)-\eps
\right]\underset{n\to\infty}{\longrightarrow }0
\end{split}
\end{equation*}
exponentially fast by the law of large numbers. For the calculation of $\bb{E}_{x^n}[P_{e,k}^{(2)}]$ we have to distinguish between the receiving nodes. We present the analysis for $k=1$, the case $k=2$ follows accordingly. 
Thereby, we use the fact that for $v=[w_1,w_2]\neq[w_1,\hat{w}_2]$ 
the random variables $p(y_1^n|X^n(v))$ and $d(X^n(w_1,\hat{w}_2),y_1^n)$ are independent for each choice of $y_1^n\in\set{Y}_1^n$.
\begin{equation*}
\begin{split}
&\bb{E}_{x^n}[P_{e,1}^{(2)}]=
\frac{1}{|\set{W}_1|\,|\set{W}_2|}\sum_{v\in\set{W}_1\times\set{W}_2}
\bb{E}_{x^n}[P_{e,1}^{(2)}(v)]\\
&\underset{\text{fixed } v}{\overset{\text{for any}}{=}}
\sum\limits_{y_1^n\in\set{Y}^n_1}
\!\!\!\bb{E}_{x^n}\!\Big[
p(y_1^n|x^n(v))\!\!\!
\sum\limits_{\substack{\hat{w}_2=1\\\hat{w}_2\neq w_2}}^{|\set{W}_2|}
\big(1-d(x^n(w_1,\hat{w}_2),y_1^n)\big)
\Big]\\
&=\sum\limits_{y_1^n\in\set{Y}^n_1}\!\!
\sum\limits_{\substack{\hat{w}_2=1\\\hat{w}_2\neq w_2}}^{|\set{W}_2|}\!\!
\bb{E}_{x^n}\!\!\big[p(y_1^n|x^n(v))\big]
\bb{E}_{x^n}\!\!\big[1-d(x^n(w_1,\hat{w}_2),y_1^n)
\big]\\
&\;=\sum\limits_{y_1^n\in\set{Y}^n_1}
\sum\limits_{\substack{\hat{w}_2=1\\\hat{w}_2\neq w_2}}^{|\set{W}_2|}
p(y_1^n)
\bb{E}_{x^n}\big[1-d(x^n(w_1,\hat{w}_2),y_1^n)
\big]
\end{split}
\end{equation*}
\begin{equation*}
\begin{split}
&\;=\sum\limits_{y_1^n\in\set{Y}^n_1}
\sum\limits_{\substack{\hat{w}_2=1\\\hat{w}_2\neq w_2}}^{|\set{W}_2|}
p(y_1^n)
\sum\limits_{x^n\in\set{X}^n}p(x^n)
\big(1-d(x^n,y_1^n)
\big)\\
&\;=(|\set{W}_2|-1)
\sum\limits_{y_1^n\in\set{Y}^n_1}
\sum\limits_{x^n\in\set{S}(y_1^n)}p(x^n)
p(y_1^n).
\end{split}
\end{equation*}
Whenever $x^n\in\set{S}(y_1^n)$, we have 
$I(x^n;y_1^n)=
\frac{1}{n}\log_2\frac{p(y_1^n|x^n)}{p(y_1^n)}
>\frac{1}{2}(R_{\overrightarrow{\mathrm{R 1}}}+I(X;Y_1))$ or
$
p(y_1^n)<p(y_1^n|x^n)2^{-n(\Rro+I(X;Y_1))/2}$. Consequently, 
\begin{equation*}
\begin{split}
\bb{E}_{x^n}[P_{e,1}^{(2)}]&<|\set{W}_2|\sum\limits_{y_1^n\in\set{Y}^n_1}
\sum\limits_{x^n\in\set{S}(y_1^n)}p(x^n)p(y_1^n|x^n)2^{-\frac{n}{2}(\Rro+I(X;Y_1))}\\
&\leq2^{n\Rro}2^{{-n}(\frac{1}{2}\Rro+\frac{1}{2}I(X;Y_1))}
=2^{{n}(\frac{1}{2}\Rro-\frac{1}{2}I(X;Y_1))}\leq2^{-n\eps}\underset{n\to\infty}{\longrightarrow }0
\end{split}
\end{equation*}
Hence, if $R_{\overrightarrow{\mathrm{R} k}}< I(X,Y_k)$, $k=1,2$, 
the average probability of error, averaged over codebooks and codewords, gets arbitrary small for sufficiently large block length $n$.

\subsubsection{Code Construction with arbitrary small maximum probability of error}\label{aver-to-max}
If $\Rro<I(X;Y_1)$ and $\Rrt<I(X;Y_2)$ we can choose 
$\eps>0$ and $n\in\N$ so that we have 
$\bb{E}_{x^n}[\mu_1^{(n)}+\mu_2^{(n)}]<\eps$. Since the average probabilities 
of error over the codebooks is small, there exists at least one codebook $\set{C}^{\star}$ with 
a small average probabilities of error $\mu_1^{(n)}+\mu_2^{(n)}<\eps$. 
This implies that we have $\mu_1^{(n)}<\eps$  and $\mu_2^{(n)}<\eps$.
We define sets 
\begin{align*}
\set{Q}&:=\{{v}\in\set{V}:\lambda_1({v})<8\eps \text{ and } \lambda_2({v})<8\eps\}\\
\set{R}_k&:=\{{v}\in\set{V}:\lambda_k({v})\geq 8\eps\} \qquad\text { for }k=1,2.
\end{align*}
Since $\eps > \frac{1}{|\set{V}|}\sum_{{v}\in\set{V}}\lambda_k({v})
\geq\frac{|\set{R}_k|}{|\set{V}|}8\eps$, we can bound the cardinality $|\set{R}_k|<\frac{|\set{V}|}{8}$ for $k=1,2$.
Then from $\set{V}=\set{Q}\cup\set{R}_1\cup\set{R}_2$ it follows
$$|\set{Q}|\geq|\set{V}|-|\set{R}_1|-|\set{R}_2|>\tfrac{3}{4}|\set{V}|.$$

Now, let $\set{T}$ be the set of $w_1$ having the property that for each $w_1$ there are at least $\frac{1}{2}M^{(n)}_2$ choices of $w_2$ so that $[w_1,w_2]\in\set{Q}$. Therefore, for $w_1\in\set{T}$ there are at most $M^{(n)}_2$ choices $w_2\in\set{W}_2$ and for $w_1\notin\set{T}$ there are less than $\frac{1}{2}{M^{(n)}_2}$ choices $w_2\in\set{W}_2$ such that $[w_1,w_2]\in\set{Q}$. Accordingly, we have
$$
|\set{T}|M^{(n)}_2+|\set{W}_1\setminus \set{T}|\tfrac{1}{2}{M^{(n)}_2}>|\set{Q}|>\tfrac{3}{4}M^{(n)}_1 M^{(n)}_2
$$
so that it follows that $|\set{T}|>\frac{1}{2}{M^{(n)}_1}$ using $|\set{W}_1\setminus \set{T}|=M^{(n)}_1-|\set{T}|$. This means that there exists an index set $\set{Q}_1^\star\subset\set{W}_1$ with $\frac{1}{2}{M^{(n)}_1}$ indices $w_1$, to each of which we can find an index set $\set{Q}_2^\star(w_1)\subset\set{W}_2$ with 
$\frac{1}{2}{M^{(n)}_2}$ indices $w_2$ so that we have for each 
$w_1\in\set{Q}_1^\star$ and $w_2\in\set{Q}_2^{\star}(w_1)$
a maximum error $\lambda_k(w_1,w_2)<8\eps$, $k=1,2$.

It follows that there exist one-to-one mappings $\Phi:\set{V}^{\star}\to\set{Q}^\star$, 
$\Phi_1:\set{W}_1^{\star}\to\set{Q}_1^\star$, $\Phi_2^{w_1}:\set{W}_2^{\star}\to\set{Q}_2^\star(w_1)$ for each $w_1\in\set{Q}_1^\star$ with $\Phi(w_1,w_2):=[\Phi_1(w_1),\Phi_2^{w_1}(w_2)]$ 
with sets $\set{V}^\star:=\set{W}_1^\star\times\set{W}_2^\star$, $\set{W}_k^\star:=\{1,2\dots,\frac{1}{2}M^{(n)}_k\}$ for $k=1,2$, 
$\set{Q}^\star:=\{[w_1,w_2]\in\set{V}:w_1\in\set{Q}_1^\star,w_2\in\set{Q}_2^\star(w_1)\}\subset\set{Q}$. 
Accordingly, there exist mappings
$\Psi_k:\set{Q}^\star\to\set{W}_k^\star$, $k=1,2$, with ${v}=[\Psi_1(\Phi({v})),\Psi_2(\Phi({v}))]$.

This allows us finally to define a $(\frac{1}{2}{M_1^{(n)}},\frac{1}{2}{M_2^{(n)}},n)$-code with an encoder $\tilde{{x}}^n:\set{V}^\star\to\set{X}^n$ with $\tilde{x}^n({v}):=x^n(\Phi({v}))$ and decoders
$\tilde{g}_1:\set{Y}^n_1\times\set{W}_1^\star\to\set{W}_2^\star$ and 
$\tilde{g}_2:\set{Y}^n_2\times\set{W}_2^\star\to\set{W}_1^\star$ with $\tilde{g}_1(y_1^n,w_1):=\tilde{\Psi}_2(w_1,g_1(y_1^n,w_1))$ and
$\tilde{g}_2(y_2^n,w_2):=\tilde{\Psi}_1(g_2(y_2^n,w_2),w_2)$ 
where we use the mappings
$\tilde{\Psi}_k:\set{V}\to W_k^\star$ given by 
$$
\tilde{\Psi}_k(v):=
\begin{cases}\Psi_k({v}),&\text{if }{v}\in\set{Q}^\star\\ 
0,&\text{if }{v}\notin\set{Q}^\star
\end{cases}
$$ 
for $k=1,2$. The idea is that the encoder uses only codewords $x^n({v})$ of the code $\set{C}^\star$ with an index ${v}\in\set{Q}^\star$, which have a maximum error $\lambda_k({v})<8\eps$, $k=1,2$. Since the decoders use the typical set decoder of the code $\set{C}^\star$, they could erroneously find an $x^n({v})$ with ${v}\in\set{V}\setminus\set{Q}^{\star}$. In this case, the mapping $\tilde{\Psi}_k$ decides on the erasure symbol $0$. 
It was already a wrong decision by the decoder $g_k$, since the encoder chooses only codewords $x^n({v})$ with ${v}\in\set{Q}^\star$. Therefore, this does not add any error to the decoding. 
The code has a rate pair $[\lfloor n(\Rrt-\frac{1}{n})\rfloor,\lfloor n(\Rro-\frac{1}{n})\rfloor]$, which can be made arbitrary close to $[\Rrt,\Rro]$ when $n\to\infty$. This proves the achievability of any rate pair satisfying the equation \eqref{eq:RATEwithoutTSH}.

\subsubsection{ Convex hull}
Let $\set{R}(p(x))$ denote the set of rates which we can achieve with the input distribution $p(x)$. Since the cardinality of the input set $\set{X}$ is finite, the rate region $\bigcup_{p(x)}\set{R}(p(x))$ is bounded. 


For $k=1,2$, we can rewrite the right hand side of  \eqref{eq:RATEwithTSH} as follows
$$I(X;Y_k|U)={\textstyle \sum\limits_{u=1}^{|\set{U}|}} p(u)I(X;Y_k|U=u)={\textstyle\sum\limits_{u=1}^{|\set{U}|}} p(u)I(X;Y_{k})\big|_{p(x|u)},$$
where in $I(X;Y_{k})\big|_{p(x|u)}$ we choose a specific input distribution $p(x|u)$ according to the auxiliary random variable $U$. For the input distribution $p(x|u)$ we know from the first part of the proof that any rate pair $\vec{R}_u
\in\set{R}(p(x|u))\subset\bb{R}^2$ is achievable. 
Therefore, for any convex combination $\sum_{u=1}^m \alpha_u\vec{R}_u$ 
we can regard the weights as probability mass function with $p(u):=\alpha_u$ and $u\in\set{U}:=\{1,2,\dots,m\}$ and choose for any $u$ an input distribution $p(x|u)$ that achieves the rate pair $\vec{R}_u$. For that reason,
the conditional mutual informations given by the right hand sides of \eqref{eq:RATEwithTSH} are also achievable rates. 
 \endproof

The coding theorem usually offers a hint how to design a good channel code practically. Accordingly, in \cite{ScOeStCISS} an interesting coset coding strategy for symmetric channels is discussed. 

In general in multi-terminal system the average and maximal error capacity region 
can be different. 
Ahlswede has shown for the two-way channel in \cite{AhliTwoWay} 
that ``one cannot reduce a code with average errors to a code with maximal 
errors without an essential loss in code length or error probability, 
 whereas for one-way channels it is unessential whether one uses average or maximal errors.''
The problem in the two-way channel is to find a maximal error sub-code with a Cartesian product structure. 
This problem is equivalent to a combinatorial problem by Zarankiewicz and
arises since the transmitter and receiver have partial knowledge only. 
Here, the relay node has full knowledge so that for the code construction with arbitrarily small 
maximum probability of error we need not require a sub-code with Cartesian product structure.

In the next subsection we prove the weak converse for the maximal error. 
Since the Fano's inequalities apply for the average error as well, the weak converse for the 
average error follows analogously.

\subsection{ Proof of weak converse} 
We have to show that  any given sequence of $(M_1^{(n)},M_2^{(n)},n)$-codes with 
$\lambda_1^{(n)},\lambda_2^{(n)}\to 0$  must satisfy $\frac{1}{n}\log M_1^{(n)}\leq I(X;Y_2|U)+o(n^0)$ and
$\frac{1}{n}\log M_2^{(n)}\leq I(X;Y_1|U)+o(n^{0})$ for a joint probability distribution $q_1(u) q_2(x|u)$ $p(y_1,y_2|x)$.
 For a  fixed block length $n$  we define the joint probability distribution $p(w_1,w_2,x^n,y_1^n,y_2^n):= \frac{1}{|\set{W}_1|} \frac{1}{|\set{W}_2|}$ $q_2(x^n|w_1,w_2) \prod_{i=1}^{n}p(y_{1i},y_{2i}|x_i)$
on $\set{W}_1\times\set{W}_2\times\set{X}^n\times\set{Y}_1^n\times\set{Y}_2^n$ where the conditional distribution $q_2(x^n|w_1,w_2)=1$ if $x^n$ is the codeword corresponding to $w_1,w_2$ or is equal to $0$ else. In what follows we consider for $k=1,2$ uniformly distributed random variables $W_k$ with values in $\set{W}_k$.

\begin{lemma}
\label{lemma:fano} For our context we have the 
Fano's inequality
\begin{equation}
\begin{split}
H(W_2|Y_1^n,W_1)\leq\lambda_1^{(n)}\log|\set{W}_2| +1=n\eps_1^{(n)},\\
\end{split}
\end{equation}
with $\eps_1^{(n)}=\frac{\log|\set{W}_2|}{n}\lambda_1^{(n)}+\frac{1}{n}\to0$ for $n\to\infty$ 
as $\lambda_1^{(n)}\to 0$.
\end{lemma}
\begin{proof}
From $Y_1^n$ and $W_1$ node 1 estimates the index $W_2$ from the sent codeword $X^n(W_1,W_2)$. We define the event of an error at node 1
as 
$$
E_1:=\begin{cases}
1, & \text{if } g_1(Y_1^n,W_1)\neq {W}_2,\\
0, & \text{if } g_1(Y_1^n,W_1)= W_2,

\end{cases}
$$ so that we have for the mean probability of error $\mu_1^{(n)}=\bb{P}[E_1=1]\leq\lambda_1^{(n)}$. 
From the chain rule for entropies we have
\begin{align}
H(E_1,W_2|Y_1^n,W_1)
&=H(W_2|Y_1^n,W_1)+H(E_1|Y_1^n,W_1,W_2)\nonumber\\
&=H(E_1|Y_1^n,W_1)+H(W_2|E,Y_1^n,W_1)\nonumber
\end{align}

Since $E_1$ is a function of $W_1,W_2$ and $Y_1^n$, we have $H(E_1|Y_1^n,W_1,W_2)=0$.
Further, since $E_1$ is a binary-valued random variable, we get $H(E_1|Y_1^n,W_1)\leq H(E_1) \leq 1$. 
So that finally with the next inequality
\begin{align}
H(W_2|Y_1^n,W_1,E_1)&
=\bb{P}[E_1=0]H(W_2|Y_1^n,W_1,E_1=0)
+\bb{P}[E_1=1]H(W_2|Y_1^n,W_1,E_1=1)\nonumber\\
&\leq(1-\mu_1^{(n)})0+\mu_1^{(n)}\log(|\set{W}_2|-1)
\leq\lambda_1^{(n)} \log|\set{W}_2|\nonumber
\end{align}
we get Fano's inequality for our context.
\end{proof}

Therewith, we can bound the entropy $H(W_2)$ as follows
\begin{align}
H(W_2)&=H(W_2|W_1)=I(W_2;Y_1^n|W_1)+H(W_2|Y_1^n,W_1)\nonumber\\
&\leq I(W_2;Y_1^n|W_1)+n\eps_1^{(n)} \leq I(W_1,W_2;Y^n_1)+n\eps_1^{(n)}\nonumber\\
&\leq I(X^n;Y^n_1)+n \eps_1^{(n)}\leq H(Y_1^n)-H(Y^n_1|X^n)+n \eps_1^{(n)}\nonumber
\end{align}
where the equations and inequalities follow from the independence of $W_1$ and $W_2$, the definition of mutual information, Lemma 1,  the chain rule for mutual information,  the positivity of mutual information, and  the data processing inequality.
If we divide the inequality by $n$ we get the rate
\begin{align}
\tfrac{1}{n}H(W_2)&\leq
\tfrac{1}{n}{\textstyle \sum\limits_{i=1}^n} \big(H(Y_{1i}|Y_1^{i-1})-H(Y_{1i}|Y_1^{i-1},X^n)\big)+\eps_1^{(n)}\nonumber\\
&\!\leq \tfrac{1}{n}{\textstyle \sum\limits_{i=1}^n} \big( H(Y_{1i})\!- \!H(Y_{1i}|X_i)\big)\!+ \!\eps_1^{(n)}
=\tfrac{1}{n}{\textstyle \sum\limits_{i=1}^n} I(Y_{1i};X_i)+\eps_1^{(n)}\nonumber
\end{align}
using the memoryless property and again standard arguments. A similar derivation for the source rate $\frac{1}{n}H(W_1)$ gives us the bound
$\tfrac{1}{n}H(W_1) \leq\tfrac{1}{n}{\textstyle \sum\limits_{i=1}^n} I(Y_{2i};X_i)+\eps_2^{(n)}
$ with $\eps_2^{(n)}=\frac{\log|\set{W}_1|}{n}\lambda_2^{(n)}+\frac{1}{n}\to0$ for $n\to \infty$ as $\lambda_2^{(n)}\to 0$.

This means that the entropies $H(W_1)$ and $H(W_2)$ are bounded by averages of the mutual informations calculated at the empirical distribution in column $i$ of the codebook. Therefore, we can rewrite these inequalities with an auxiliary random variable
$U$, where $U=i\in\set{U}=\{1,2,\dots,n\}$ with probability $\frac{1}{n}$. We finish the proof of the converse with the following inequalities
\begin{align*}
\tfrac{1}{n}H(W_2) 
&\leq  \tfrac{1}{n}{\textstyle \sum\limits_{i=1}^n}
I(Y_{1i};X_i)+\eps_1^{(n)}\nonumber\\
&={\textstyle \sum\limits_{i=1}^n} \bb{P}(U=i)I(Y_{1i};X_i|U=i)+\eps_1^{(n)}\nonumber\\
&= I(Y_{1U};X_U|U) +\eps_1^{(n)} = I(Y_1;X|U) +\eps_1^{(n)}\label{eq:empI}
\end{align*}
and  $\tfrac{1}{n}H(W_1)\leq I(Y_2;X|U) +\eps_2^{(n)}$ accordingly where $\eps_k^{(n)}\to 0$, $k=1,2$, when $n\to\infty$. Thereby, $Y_k:=Y_{kU}$ and $X:=X_U$ are new random variables whose distribution depend on $U$ in the same way as the distributions of $Y_{ki}$ and $X_i$ depend on $i$. 
\endproof

Up to now the auxiliary random variable $U$ is defined on a set $\set{U}$ with arbitrary cardinality. 
Next, we will show that  $|\set{U}|=2$ is enough.


\subsection{Cardinality of set $\set{U}$ }
\label{subsect:card}

With Fenchel--Bunt's extension of Carath\'eodory's theorem it follows that any rate pair in 
$\text{ConvexHull}\big(\bigcup_{p(x)}\set{R}(p(x))\big)=\bigcup_{u\in\set{U}}\set{R}(p(x|u))$ 
is achievable by time-sharing between two rate pairs
from $\bigcup_{p(x)}\set{R}(p(x))$, i.e. $|\set{U}|=2$ is enough. 

\begin{theorem}[{\cite[Theorem 1.3.7]{HiriartLemar}}]
\label{theo:fenchel}
If $\set{S}\subset\bb{R}^n$ has no more than $n$ connected components (in particular, if $\set{S}$ is connected), then any $x\in\mathrm{ConvexHull}(\set{S})$ can be expressed as a convex combination of $n$ elements of $\set{S}$.
\end{theorem}

Since for any $x\in\set{X}$ we have $[0,0]\in\set{R}(p(x))$, the set $\bigcup_{p(x)}\set{R}(p(x))$ is connected. Therefore,
any rate pair in $\set{C}_{\mathrm{BC}}=\text{ConvexHull}\big(\bigcup_{p(x)}\set{R}(p(x))\big)$ can be expressed as a convex combination of $n=2$ rate pairs of $\bigcup_{p(x)}\set{R}(p(x))$. 
\endproof

This finishes the proof of the capacity region of the bidirectional broadcast channel.
\begin{remark}
Since the coding theorem includes the achievability of rate pairs in terms of the average probability of error and the proof of the weak converse for the average error works analogously, $\sCbc$ is also the  capacity region in terms of average probability of error. 
\end{remark}

\begin{remark}
The characterization of the bidirectional broadcast capacity region for Gaussian channels is analogous. We would have to deal with discrete channels with Gaussian channel transfer distributions and would  have to add an input power constraints but the arguments are similar to the arguments considered here. 
\end{remark}

In the next section we present the strong converse in the case of maximum probability error. Therefore, we will refine the achievability definition to $[\eps_1,\eps_2]$-achievable rate pairs. Then it follows from the strong converse for the maximum probability of error that the $[\eps_1,\eps_2]$-capacity region is equal $\sCbc$. Finally, from this we can deduce on the $[\eps_1,\eps_2]$-capacity region in terms of average probability of error for sufficiently small average error. 


\section{Sharper Versions of the Converse Part for the Broadcast Phase}

\label{sec:StrongConv}
Here, we derive a sharper converse to the coding theorem for the bidirectional broadcast channel. We prove the full strong converse for the capacity region defined with respect to the maximum error probability, i.e. $\sCbcmax(\eps_1,\eps_2)= \sCbc$ for all $\eps_1,\eps_2\in (0,1)$. Additionally, we show that the $[\eps_1,\eps_2]$-capacity region $\sCbcav(\eps_1,\eps_2)$ defined by using average error probability coincides with $\sCbc$ for \emph{small values} of $\eps_1,\eps_2\in (0,1)$.

 The main tool we will use is the powerful blowing-up technique introduced by Ahlswede, G\'acs and K\"orner in \cite{AGK} based on the Blowing-up Lemma (cf. Marton's  paper \cite{martonBlowUp} for a simpler information-theoretic proof). The basic idea developed in \cite{AGK} is that blowing-up the decoding sets in conjunction with a variant of Fano's inequality allows us to convert the weak converse into the strong converse to the coding theorem.
 
Before entering the proof we recall the essential blowing-up notations and results which we need in the sequel: For a finite set $\mathcal{Y}$, $n,l\in\N$ and $\set{B}\subset \mathcal{Y}^{n} $ we define the \emph{Hamming l-neighborhood} by
\[ \Gamma^{l}\set{\set{B}}:=\{y\in \mathcal{Y}^n: d_{H}(y,\set{B})\le l  \},\]
where $d_{H}$ denotes the non-normalized Hamming metric and $d_{H}(y;\set{B}):=\min_{y'\in \set{B}}d_{H}(y,y')$.
\begin{theorem}[Ahlswede/G\'acs/K\"orner \cite{AGK}, cf. also \cite{CsiszarKoerner}, \cite{martonBlowUp}]\label{blowing-up}
Let $\mathcal{X}$ and $\mathcal{Y}$ be finite sets.
\begin{enumerate}
\item For any sequence of positive integers $\{l_n\}_{n\in\N}$ with $\lim_{n\to\infty}\frac{l_n}{n}=0$ there exists a sequence $\{\delta_n(l_n,|\mathcal{Y}| )\}_{n\in \N}$ with $\lim_{n\to\infty}\delta_{n}(l_n,|\mathcal{Y}| )=0$ such that for any $\set{B}\subset \mathcal{Y}^{n}$
\[ |\Gamma^{l_n}\set{B}|\le |\set{B}|2^{n\delta_n(l_n,|\mathcal{Y}| )}. \]
\item (Blowing-Up Lemma) To any sequence $\{\eta_n\}_{n\in\N}$ with $\lim_{n\to\infty}\eta_n=0$ there exist a sequence of positive integers $\{l_n\}_{n\in \N}$ with $\lim_{n\to\infty}\frac{l_n}{n}=0$ and a sequence $\{\eps_n\}_{n\in\N}$ with $\lim_{n\to\infty}\eps_n=0$ such that for every probability transition function $p:\mathcal{X}\to \mathcal{Y}$ and every $n\in\N, x\in \mathcal{X}^n$, $\set{B}\subset \mathcal{Y}^n$  
\[p^{(n)}(\set{B}|x)\ge 2^{-n\eta_n}\textrm{ implies } p^{(n)}(\Gamma^{l_n}\set{B}|x)\ge 1-\eps_n,  \]
where $p^{(n)}:\set{X}^{n}\to\set{Y}^n$ denotes the $n$-th memoryless extension of $p$.
\end{enumerate}
\end{theorem}
\begin{remark}
The second part of Theorem \ref{blowing-up} is the \emph{uniform version} of the Blowing-up lemma according to Csiszar/K\"orner \cite{CsiszarKoerner} chap. 1.5. 
\end{remark}

Since blowing up is an operation on the subsets of the output alphabet it is convenient to describe the decoding functions $g_1$ and $g_2$ by decoding sets. This equivalent description is obtained as follows; for each fixed $w_1\in \mathcal{W}_1$ the map $g_1(\cdot ,w_1):\mathcal{Y}_{1}^{n}\to\mathcal{W}_2\cup \{0\}$ induces a partition $\mathcal{P}_{w_1}^{(n)}:=\{\set{A}_{w_2}^{'(n)}(w_1)\}_{w_2\in\mathcal{W}_2\cup \{0\}}$ of $\mathcal{Y}_{1}^{n}$. In a similar fashion for each $w_2\in \mathcal{W}_2$ we obtain, using the decoder $g_2$, a partition $\mathcal{Q}_{w_2}^{(n)}:=\{\set{B}_{w_1}^{'(n)}(w_2)\}_{w_1\in\mathcal{W}_1\cup\{0\}}$ of the output set $\mathcal{Y}_{2}^{n}$. Now if we are given the corresponding encoder $x^n:\mathcal{W}_1\times \mathcal{W}_2\to\mathcal{X}^{n}$, the probabilities of error can be expressed by
\[ \lambda_{1}(w_1,w_2)=p((\set{A}_{w_2}^{'(n)}(w_1))^{c}|x^n(w_1,w_2)),\]
and   
\[\lambda_{2}(w_1,w_2)=p((\set{B}_{w_1}^{'(n)}(w_2))^{c}|x^n(w_1,w_2)). \]
In what follows $\lambda_{k}^{(n)}$, $k=1,2$, denotes the maximum probability of error for a given code.

A pair of non-negative reals $[\Rrt,\Rro]$ is said to be $[\eps_1,\eps_2]$-achievable, $\eps_1,\eps_2\in (0,1)$, if for each $\delta>0$ there is a sequence of $(M_{1}^{(n)},M_{2}^{(n)},n)$-codes such that for all sufficiently large $n$ the following statements are fulfilled
\begin{enumerate}
\item $\frac{1}{n}\log M_{1}^{(n)}\ge \Rrt-\delta $ and $ \frac{1}{n}\log M_{2}^{(n)}\ge \Rro-\delta$.
\item $\lambda_{k}^{(n)}\le \eps_k$ for $k=1,2$.
\end{enumerate}
The set of all $[\eps_1,\eps_2]$-achievable rates with respect to the maximum probability of error is denoted by $\sCbcmax(\eps_1,\eps_2)$. It is clear that $\sCbc\subseteq \sCbcmax(\eps_1,\eps_2) $ and
\[\sCbc=\bigcap_{\eps_1,\eps_2\in (0,1)}\sCbcmax(\eps_1,\eps_2)   \]
hold. The content of the strong converse is that $\sCbc$ cannot be a proper subset of $\sCbcmax(\eps_1,\eps_2)$ for $\eps_1,\eps_2\in (0,1)$:
\begin{theorem}\label{strong-converse-max-error}
For memoryless bidirectional broadcast channel we have 
\[ \sCbc=\sCbcmax(\eps_1,\eps_2)\]
for all $\eps_1,\eps_2\in (0,1)$.
\end{theorem}
\begin{proof} Let $[\Rrt,\Rro]$ be an $[\eps_1,\eps_2]$-achievable rate pair, thus, by definition, for any $\delta>0$ we can find a sequence of $(M_{1}^{(n)},M_{2}^{(n)},n)$-codes and $n(\delta)\in\N$ such that for all $n\ge n(\delta)$ following conditions are satisfied:
 \begin{enumerate}
\item $\frac{1}{n}\log M_{1}^{(n)}\ge \Rrt-\delta $ and $ \frac{1}{n}\log M_{2}^{(n)}\ge \Rro-\delta$.
\item $\lambda_{k}^{(n)}\le \eps_k$ for $k=1,2$.
\end{enumerate}
For those $n$ we consider the families of partitions associated with the decoder maps, i.e. for each $w_1\in \mathcal{W}_1$ we have a partition $\mathcal{P}_{w_1}^{(n)}=\{\set{A}_{w_2}^{'(n)}(w_1)\}_{w_2\in\mathcal{W}_2\cup \{0\}}$ of $\mathcal{Y}_{1}^{n}$ and analogously for each $w_2\in\mathcal{W}_2$ a partition $\mathcal{Q}_{w_2}^{(n)}=\{ \set{B}_{w_1}^{'(n)}(w_2)\}_{w_1\in\mathcal{W}_1\cup \{0\}}$ of $\mathcal{Y}_{2}^{n}$ such that for all $w_1\in \set{W}_1$ and $w_2\in\set{W}_2$ we have
\[ p(\set{A}_{w_2}^{'(n)}(w_1)|x^n(w_1,w_2) )\ge 1-\eps_1\ge 2^{-n\eta_n},\]
and 
\[ p(\set{B}_{w_1}^{'(n)}(w_2)|x^n(w_1,w_2) )\ge 1-\eps_2 \ge 2^{-n\eta_n} \]
where $\eta_n:=\frac{1}{n}\max \{-\log(1-\eps_1),-\log(1-\eps_2)\}$. According to the second part of Theorem \ref{blowing-up} we can find a sequence of positive integers $\{l_n\}_{n\in\N}$ with $\lim_{n\to\infty}\frac{l_n}{n}=0$ such that for the sets
\[ \set{A}_{w_2}(w_1):=\Gamma^{l_n}\set{A}_{w_2}^{'(n)}(w_1)\textrm{ and } \set{B}_{w_1}(w_2):=\Gamma^{l_n}\set{B}_{w_1}^{'(n)}(w_2),\]
we have
\begin{equation}\label{sc-0} 
p(\set{A}_{w_2}(w_1)|x^n(w_1,w_2))\ge 1-\eps_n,
\end{equation}
and
\[ p(\set{B}_{w_1}(w_2)|x^n(w_1,w_2))\ge 1-\eps_n  \]
with $\lim_{n\to\infty}\eps_n=0$. The sets $\{\set{A}_{w_2}(w_1)\}_{w_2\in\mathcal{W}_2}$ are not necessarily disjoint for different values of $w_2$. The same applies to the sets $\{\set{B}_{w_1}(w_2)\}_{w_1\in\mathcal{W}_1}$. Nevertheless, we show now that for any given $w_1\in \mathcal{W}_1$ each $y_1^n\in \mathcal{Y}_{1}^{n}$ is contained in at most sub-exponentially many $\set{A}_{w_2}(w_1) $. To this end, for any given $y_1^n\in\mathcal{Y}_{1}^{n}$ and $w_1\in\mathcal{W}_1$ we define the set
\[ O_1(y_1^n,w_1):=\{w_2\in\mathcal{W}_2: y_1^n\in \set{A}_{w_2}(w_1)\},\]
and claim that 
\begin{equation}\label{sc-1} 
|O_{1}(y_1^n,w_1)|\le 2^{n\delta_{n}(l_n,|\mathcal{Y}|)},
\end{equation}
with $\lim_{n\to\infty}\delta_n(l_n,|\mathcal{Y}|)=0$ holds. The proof is given in \cite{AGK}. We reproduce the full argument for convenience. It is obvious that $w_2\in O_1(y_1^n,w_1)$ if and only if $\set{A}_{w_2}^{'(n)}(w_1)\cap \Gamma^{l_n}\{y_1^n\}\neq\emptyset$. Therefore, since the sets $\{\set{A}_{w_2}^{'(n)}(w_1)\}_{w_2\in\mathcal{W}_2}$ are disjoint, we have 
\[|O_{1}(y_1^n,w_1)|\le |\Gamma^{l_n}\{y_1^n\}|\le 2^{n\delta_{n}(l_n,|\mathcal{Y}|)}   \]
with $\lim_{n\to\infty}\delta_n(l_n,|\mathcal{Y} |)=0 $ by the first part of Theorem \ref{blowing-up}. A similar result holds for the analogously defined set $O_2(y_2^n,w_2)$.

Let us consider two independent, uniformly distributed random variables $W_1$ and $W_2$ taking values in the sets $\mathcal{W}_1$ and $\mathcal{W}_2$ and a random variable $X^n$ with values in $\mathcal{X}^n$ such that
\begin{equation}\label{sc-x}
\mathbb{P}(X^n=x^n(w_1,w_2)| W_{1}=w'_1, W_2=w'_2)=\delta_{(w_1,w_2),(w'_1,w'_2)}.  
\end{equation}
Then the probability distribution of the whole system is given by
\begin{equation}\label{sc-y}
\begin{split}
  p(w_1,w_2, x^n(w'_1,w'_2),y_1^n,y_2^n)=\tfrac{1}{M_{1}^{(n)}M_{2}^{(n)}}\delta_{(w_1,w_2),(w'_1,w'_2)}\; p(y^n_1,y^n_2|x^n(w'_1,w'_2 ))
\end{split}
\end{equation}
for $w_k,w'_k\in \mathcal{W}_k$, $ x^n(w'_1,w'_2)\in \mathcal{X}^n $, $y_k^n\in \mathcal{Y}_{k}^{n}$ and $k=1,2$.

Furthermore, for given $y_1^n\in \mathcal{Y}_{1}^{n}$ and $w_1\in \mathcal{W}_1$ let us define
\begin{equation}\label{sc-2}
  \eps (y_1^n,w_1):=\mathbb{P}(W_2\notin O_1(y_1^n,w_1)|Y_{1}^{n}=y_1^n,W_1=w_1 ).
\end{equation}
As in the proof of the weak converse one can show that
\begin{eqnarray}\label{sc-3}
\log M_{2}^{(n)} = H(W_2)
 \le& I(X^n;Y_{1}^{n})+H(W_2|Y_{1}^n,W_1) 
\end{eqnarray}
holds. Now, we need a variant of Fano's inequality which incorporates the quantity defined in (\ref{sc-2}). Therefore, we  use the following elementary entropy inequality: For a probability distribution $p$ on a finite set $\set{A}$ and an arbitrary $\set{B}\subseteq \set{A}$ we have
\begin{equation}\label{sc-4}
-\sum_{x\in \set{B}}p(x)\log p(x)\le -p(\set{B})\log p(\set{B})+p(\set{B})\log|\set{B}|.  
\end{equation}
Then for given $y_1$ and $w_1$ we set 
\[ H(W_2)_{y^n_1,w_1}:=H(W_2|Y_{1}^n=y^n_1, W_1=w_1)\]
and obtain
\begin{equation}\label{sc-5}
\begin{split}
H(W_2)_{y^n_1,w_1}&=
 -\sum_{w_2\in O_1(y^n_1,w_1)}p(w_2|y^n_1,w_1) \log p(w_2|y^n_1,w_1)\\
&\qquad\qquad -\sum_{w_2\notin O_1(y^n_1,w_1)}p(w_2|y^n_1,w_1) \log p(w_2|y^n_1,w_1) \\
&\le  H(\eps (y^n_1,w_1))
 +\eps (y^n_1,w_1)\log M_{2}^{(n)} 
 +(1-\eps(y^n_1,w_1))n\delta_n, 
\end{split}
\end{equation}
where we have applied eq. (\ref{sc-4}) to each sum and then used eq. (\ref{sc-1}) with the abbreviation $\delta_n=\delta_n(l_n,|\mathcal{Y}|)$. $H(\eps (y^n_1,w_1)) $ denotes the entropy of the distribution $(\eps(y^n_1,w_1), 1-\eps(y^n_1,w_1) )$. Averaging with respect to $\mathbb{P}(Y_{1}^{n}=y_1, W_1=w_1)$ and using the concavity of the entropy we arrive at
\begin{eqnarray}\label{sc-6}
  H(W_2|Y_{1}^{n},W_1)&\le& H(\tau_n)+\tau_n\log M_{2}^{(n)}
 +(1-\tau_n)n\delta_n,
\end{eqnarray}
with $\tau_n:=\sum_{y_1\in \mathcal{Y}_{1}^{n},w_1\in \mathcal{W}_1}\mathbb{P}(Y_{1}^{n}=y^n_1, W_1=w_1)\eps(y^n_1,w_1) $.
Note that by (\ref{sc-2}), our definition of $X^n$ in eq. (\ref{sc-x}) and (\ref{sc-y}) we have
\begin{eqnarray}\label{sc-7}
\tau_n&=& \sum_{w_1,y^n_1}\sum_{w_2\notin O_{1}(y^n_1,w_1)}p(w_1,w_2,y^n_1)\nonumber\\
&=& \sum_{w_1,y^n_1}\sum_{w_2\notin O_{1}(y^n_1,w_1)}\frac{1}{M_{1}^{(n)}M_{2}^{(n)}} p(y^n_1|x^n(w_1,w_2))\nonumber\\
&=&\sum_{w_1,w_2}\frac{1}{M_{1}^{(n)}M_{2}^{(n)}}p((\set{A}_{w_2}(w_1))^c|x^n(w_1,w_2))\nonumber\\
 &\le& \eps_n,
\end{eqnarray}
where the third equality holds since $w_2\notin O_{1}(y^n_1,w_1)$ iff $y^n_1\notin \set{A}_{w_2}(w_1)$ and the last inequality is by eq. (\ref{sc-0}).
Thus (\ref{sc-3}), (\ref{sc-6}) and (\ref{sc-7}) show that
\begin{equation}\label{sc-8}
  \frac{1}{n}\log M_{2}^{(n)}\le \frac{1}{n}I(X^n;Y_{1}^{n})+o(n^{0}).
\end{equation}
Similar reasoning shows that
\begin{equation}\label{sc-9}
   \frac{1}{n}\log M_{1}^{(n)}\le \frac{1}{n}I(X^n;Y_{2}^{n})+o(n^{0})
\end{equation}
also holds. It is obvious that as in the proof of the weak converse the mutual informations on the right hand sides of (\ref{sc-8}) and (\ref{sc-9}) can be written as $I(X;Y_1|U_n)$ and $I(X;Y_2|U_n)$ for a suitable random variable $U_n$ taking values in $\{1,\ldots ,n\}$. Note that by the proof of the coding theorem with the weak converse the rates $I(X;Y_1|U_n) $ and $I(X;Y_2|U_n) $ are achievable. Thus, we can conclude our proof by noting that for sufficiently large $n$ we have
\[ \Rro-\delta\le \frac{1}{n}\log M_{2}^{(n)}\le \frac{1}{n}I(X^n;Y_{1}^{n})+o(n^{0}) ,  \]
and 
\[ \Rrt-\delta \le  \frac{1}{n}\log M_{1}^{(n)}\le \frac{1}{n}I(X^n;Y_{2}^{n})+o(n^{0}), \]
and that $\sCbc$ is closed. This shows that $\sCbcmax(\eps_1,\eps_2)\subset \sCbc$ and we are done.
\end{proof}
We give now the partial extension of Theorem \ref{strong-converse-max-error} to the capacity region $\sCbcav(\eps_1,\eps_2)$ which is defined similarly to $\sCbcmax(\eps_1,\eps_2)$ the difference being only that we use the average probability of error. Our strategy will be to reduce the statement to the Theorem \ref{strong-converse-max-error} for sufficiently small $\eps_1,\eps_2\in (0,1)$.
\begin{corollary}
For memoryless bidirectional broadcast channel it holds that
\[ \sCbc=\sCbcav(\eps_1,\eps_2)   \]
for all $\eps_1\in (0,\frac{1}{2})$ and $\eps_2\in (0,\frac{1}{4})$ or $\eps_1\in (0,\frac{1}{4})$ and $\eps_2\in (0,\frac{1}{2})$.
\end{corollary}
\begin{proof}
Let $[\Rrt,\Rro]\in \sCbcav(\eps_1,\eps_2)$ with $\eps_1\in (0,\frac{1}{2})$ and $\eps_2\in (0,\frac{1}{4})$. Thus, for each $\delta>0$ there is a sequence of $(M_{1}^{(n)},M_{2}^{(n)},n) $-codes and $n(\delta)\in \N$ with
 \begin{enumerate}
\item $\frac{1}{n}\log M_{1}^{(n)}\ge \Rrt-\delta $ and $ \frac{1}{n}\log M_{2}^{(n)}\ge \Rro-\delta$.
\item $\mu_{k}^{(n)}\le \eps_k$ for $k=1,2$,
\end{enumerate}
for all $n\ge n(\delta)$ where $\mu_{k}^{(n)}$ denotes the average error probability. It is clear that $\frac{\eps}{1-\eps}<1$ iff $\eps\in (0,\frac{1}{2})$ and $3 \frac{\eps}{1-\eps}<1$ iff $\eps\in (0,\frac{1}{4})$. Therefore, we can choose real numbers $a_1\in (0,1)$ with $\frac{\eps_1}{1-\eps_1}<a_1<1 $ and $a_2\in (0,1)$ with $3\frac{\eps_2}{1-\eps_2}<a_2<1 $. Let us consider the reals $f_k(\eps_k):=\eps_k +a_k(1-\eps_k)=(1-a_k)\eps_k +a_k\in (0,1)$ for $k=1,2$.

If we define the sets
\[\mathcal{R}_k:=\{ v\in \mathcal{W}_1\times \mathcal{W}_2: \lambda_{k}(v)\ge f_k(\eps_k)\},\]
for $k=1,2$, from the Markov's inequality it is clear that 
\begin{equation}\label{sc-av-1}
  |\mathcal{R}_k|\le e_k(\eps_k)M_{1}^{(n)}M_{2}^{(n)}
\end{equation}
with $e_k(\eps_k):=\frac{\eps_k}{f_k(\eps_k)}$ for $k=1,2$.
For the set $\mathcal{Q}:=(\mathcal{R}_1\cup \mathcal{R}_2  )^{c}$ we obtain the following cardinality bound by (\ref{sc-av-1}):
\begin{equation}\label{sc-av-2}
  |\mathcal{Q}|\ge (1-\sum_{k=1}^{2}e_k(\eps_k))M_{1}^{(n)}M_{2}^{(n)}.
\end{equation}
Let 
\begin{equation}\label{sc-av-t}
\begin{split}
\mathcal{T}:=\{w_1\in \mathcal{W}_1: \textrm{there are at least }e_2(\eps_2)M_{2}^{(n)} w_2\in\mathcal{W}_2  
\textrm{ with }(w_1,w_2)\in \mathcal{Q} \}.
\end{split}  
\end{equation}
Our goal now is to find a lower bound on the cardinality of $\mathcal{T}$. To this end, note that for $w_1\in\mathcal{T}$ there are at most $M_{2}^{(n)}$ message indices $w_2\in \mathcal{W}_2$ with $(w_1,w_2)\in \mathcal{Q}$ and for $w_1\notin \mathcal{T}$ there are at most $e_2(\eps_2)M_{2}^{(n)}$ message indices $w_2\in\mathcal{W}_2$ with $(w_1,w_2)\in \mathcal{Q}$. Thus by (\ref{sc-av-2}) 
\[ 
\begin{split}
(1-{\textstyle\sum\limits_{k=1}^{2}}e_k(\eps_k))M_{1}^{(n)}M_{2}^{(n)}\le |\mathcal{Q}|\le M_{2}^{(n)}|\mathcal{T}| + e_2(\eps_2)M_{2}^{(n)} |\mathcal{W}_1\setminus \mathcal{T}|,
\end{split}  
\]
and therefore
\begin{equation}\label{sc-av-3}
  |\mathcal{T}|\ge  \frac{1-e_1(\eps_1)-2e_2(\eps_2)}{1-e_2(\eps_2)}M_{1}^{(n)}.
\end{equation}
The first factor on the right hand side of (\ref{sc-av-3}) is positive due to our restriction to $\eps_1\in (0,\frac{1}{2})$ and $\eps_2\in(0,\frac{1}{4})$. Indeed, it is easily seen that $e_1(\eps_1)<\frac{1}{2}$ iff $\frac{\eps_1}{1-\eps_1}<a_1$ and this last relation is true by our choice of $a_1$ which is possible due to our restriction to $\eps_1\in (0,\frac{1}{2})$. Similarly, we have $e_2(\eps_2)<\frac{1}{4}$ iff $3\frac{\eps_2}{1-\eps_2}<a_2$ which is satisfied since $\eps_2\in (0,\frac{1}{4})$.

Now we set 
\[ N_{1}^{(n)}:=\left\lceil \frac{1-e_1(\eps_1)-2e_2(\eps_2)}{1-e_2(\eps_2)}M_{1}^{(n)}  \right\rceil\]
and 
\[ N_{2}^{(n)}:=\left\lceil e_2(\eps_2)M_{2}^{(n)}  \right\rceil. \]
As in the proof of the direct part of the coding theorem we can construct a sequence of $( N_{1}^{(n)}, N_{2}^{(n)},n )$-codes from the given sequence of codes but with the additional property that the new sequence has the \emph{maximum} error probabilities bounded by $f_1(\eps_1)$ and $f_2(\eps_2)$. The new sequence of codes achieves the rate pair $[\Rrt,\Rro]$. Thus, we can apply our Theorem \ref{strong-converse-max-error} to conclude that for $\eps_1\in (0,\frac{1}{2})$ and $\eps_2\in (0,\frac{1}{4}) $ $[\Rrt,\Rro]\in \sCbc$. If we interchange the roles of $\mathcal{W}_1$ and $\mathcal{W}_2$ in definition of the set $\mathcal{T}$ in (\ref{sc-av-t}) and at the same time swap the numbers $f_1(\eps_1)$ and $f_2(\eps_2)$, we can conclude in a similar fashion that $ \sCbc=\sCbcav(\eps_1,\eps_2) $ for $\eps_1\in (0,\frac{1}{4})$ and $\eps_2\in (0,\frac{1}{2})$.
\end{proof}


\section{Discussion}
\label{sec:discussion}
The coding principles of the bidirectional broadcast are similar to the network coding approach where we would have implemented a bitwise XOR operation on the decoded messages at the relay node  \cite{WuChouKung05}, \cite{FBW06nca}. But since network coding \cite{Ahli}, \cite{NOWYeung} is  originally a multi-terminal source coding problem, the achievable rates in the broadcast phase using the network coding approach are limited by the worst receiver. This means that with a network coding approach we can achieve $$\Rrt,\Rro\leq\min\{I(X;Y_1),I(X;Y_2)\}$$ 
for some common input distribution $p(x)$. The achievable rates depend on the common input distribution and \emph{both} channel transfer distributions.
For our coding scheme each achievable rate depends on 
the common input distribution and its own channel transfer distribution only. For each channel we can separately find the optimal input distribution which achieves the maximal achievable rate for this link (equal to the single link capacity), but the optimal input distribution for one channel needs not  be optimal for the other channel.\footnote{It is curious that if we transfer the result to scalar Gaussian channels with a mean power constraint, obviously the Gaussian input distribution will maximize both links simultaneously. For the vector valued Gaussian channel this is no longer the case.}

Accordingly, we see that the network coding approach using XOR on the decoded messages at the relay is in general inferior, but it achieves the capacity of the bidirectional broadcast if and only if for the maximizing input distribution $p^\star(x)={\arg\max}_{p(x)}\max\{I(X;Y_1),I(X;Y_2)\}$
we have $I(X;Y_1)=I(X;Y_2)$. 

In the following we will discuss the bidirectional broadcast 
for a binary symmetric broadcast channel and the achievable rate region 
of two-phase bidirectional relaying protocol.


\subsection{Binary Symmetric Broadcast Channel}

For the binary symmetric broadcast channel, let $p_1$ and $p_2$ denote the probability that a relay input $X\in\{0,1\}$ is complemented at the output $Y_1\in\{0,1\}$ and $Y_2\in\{0,1\}$ of node 1 and 2 respectively. From \cite[Chapter 8.1.4]{CoverThomas} we know that a uniform input distribution maximizes the binary symmetric channel. Therefore, the broadcast capacity region for the binary symmetric channel is given by
\begin{equation}
\sCbc=[0,1-H(p_2)]\times [0,1-H(p_1)],
\label{eq:bsbc}
\end{equation}
which includes the region $[0,1-\max\{H(p_1),H(p_2)\}]\times[0,1-\max\{H(p_1),H(p_2)\}]$ achievable using XOR at the relay node according to \cite{WuChouKung05}.


\subsection{Achievable Bidirectional Rate Region}
We will now look at the achievable bidirectional rate region where we use in each phase the optimal strategies. Thereby, we optimize the time-division between the 
MAC phase with memoryless multiple access channel $p(y|x_1,x_2)$ and BC phase with memoryless broadcast channel $p(y_1,y_2|x)$. Of course, due to the a priori separation into two phases, this strategy need not be the optimal strategy for the bidirectional relay channel.

Let $R_1$ and $R_2$ denote the achievable rates for transmitting a messages $w_1$ from node 1 to node 2 and a message $w_2$ from node 2 to node 1 with the support of the relay node. In more detail, node 1 wants to transmit message $w_1$ with rate $nR_1$ in $n$ channel uses of the bidirectional relay channel to node 2. Simultaneously, node 2 wants to transmit message $w_2$ with rate $nR_2$ in $n$ channel uses to node 1. Then let $\nmac$ and $\nbc=n-\nmac$ denote the number of channel uses in the MAC phase and BC phase with the property $\frac{\nmac}{n}\to\alpha\in[0,1]$ and $\frac{\nbc}{n}\to 1-\alpha$ when $n\to\infty$, respectively. We call $\alpha$ the time-division factor between multiple access and broadcast phase. With a sufficient block length $n$ (respectively $\nmac$ and $\nbc$) we can achieve a bidirectional transmission of messages $w_1$ and $w_2$ with arbitrary small decoding error if rate pairs $[\Ror,\Rtr]\in\sCmac$ and $[\Rrt,\Rro]\in\sCbc$ exist so that we have
\begin{align*}
n R_1\leq \min\{\nmac\Ror,\nbc\Rrt\},\\
n R_2\leq \min\{\nmac\Rtr,\nbc\Rro\}.
\end{align*}
Thus, the achievable rate region of the bidirectional relay channel using time-division is given by the set of all rate pairs $[R_1,R_2]$ which are achievable with any time-division factor $\alpha\in[0,1]$ as $n\to\infty$. We collect the previous consideration in the following proposition.

\begin{proposition}
The achievable rate region $\sRbrc$ of the two-phase bidirectional relay channel is given by
\begin{align*}
\sRbrc=\big\{&[R_1,R_2]\in\bb{R}^2: R_1\leq \min\{\alpha \Ror,(1-\alpha) \Rrt\},\\
&R_2\leq \min\{\alpha \Rtr,(1-\alpha) \Rro\} \text{ with }\;\alpha\in (0,1),\\
&[\Ror,\Rtr]\in\sCmac,\text{ and }\;[\Rrt,\Rro]\in\sCbc\big\}.
\end{align*}
\end{proposition}

Since $\sCbc$ is larger than the region of the broadcast phase achieved  by applying interference cancellation  \cite{RW05ses,OB06oraf} and  XOR on the decoded messages at the relay node \cite{WuChouKung05,FBW06nca,NOWYeung}, the achievable rate region $\sRbrc$ includes the region which can be achieved by interference cancellation and network coding approaches.

\begin{figure}
\begin{center}	
	\includegraphics[width=8cm]{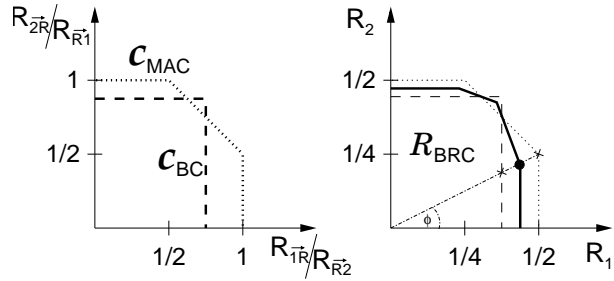}
\end{center}
\caption{The left figure shows the capacity regions $\sCmac$ (dotted line) and $\sCbc$ (dashed line), the right figure shows the corresponding achievable rate region $\sRbrc$ (solid line). The dashed-dotted line exemplarily shows for one angle $\phi$
 the achievable rate pair ($\bullet$) on the boundary of $\sRbrc$ with the optimal time-division between the two rate pairs ($\times$) on the boundary of $\sCmac$ and $\sCbc$.}
\label{fig:capreg}
\end{figure}

Finally, we briefly look at an example with binary channels. In Figure \ref{fig:capreg} we depicted the capacity region $\sCmac$ and $\sCbc$ and the achievable rate region $\sRbrc$ with a symmetric binary erasure multiple access channel \cite[Example 14.3.3]{CoverThomas} and a binary symmetric broadcast channel, cf. equation \eqref{eq:bsbc}. The boundary of the achievable rate region can be obtained geometrically if one takes for any angle $\phi\in[0,\pi/2]$ half of the arithmetical mean between the boundary rate pairs of the capacity regions where we have
$\tan \phi=\Rtr/\Ror=\Rro/\Rrt$.

 
\section{Conclusion}
In this work we present the broadcast capacity region of the two-phase bidirectional relay channel. Thereby, each receiving node has perfect knowledge about the message intended for the other node. Furthermore, the proposed achievable rate region of the two-phase bidirectional relay channel is in general larger than the rate region which can be achieved by applying the network coding principle on the decoded data. The coding theorem and weak converse are easily extended to Gaussian channels with input power constraints.

We have also shown the strong converse with respect to the maximum error criterion for the broadcast phase. This result implies then that the capacity region defined with respect to the average error probability remains constant for all error parameters $[\eps_1,\eps_2]\in (0,\frac{1}{2})\times (0,\frac{1}{4})$ or $[\eps_1,\eps_2]\in (0,\frac{1}{4})\times (0,\frac{1}{2})$. 


\end{document}